\documentstyle[aps,preprint,epsfig]{revtex}
\newcommand{\bel}[1]{\begin{equation}\label{#1}}

\newcommand{\exval}[1]{\mbox{$\langle \, {#1}\, \rangle$}}
\def\bbbc{{\mathchoice {\setbox0=\hbox{$\displaystyle\rm C$}\hbox{\hbox
to0pt{\kern0.4\wd0\vrule height0.9\ht0\hss}\box0}}
{\setbox0=\hbox{$\textstyle\rm C$}\hbox{\hbox
to0pt{\kern0.4\wd0\vrule height0.9\ht0\hss}\box0}}
{\setbox0=\hbox{$\scriptstyle\rm C$}\hbox{\hbox
to0pt{\kern0.4\wd0\vrule height0.9\ht0\hss}\box0}}
{\setbox0=\hbox{$\scriptscriptstyle\rm C$}\hbox{\hbox
to0pt{\kern0.4\wd0\vrule height0.9\ht0\hss}\box0}}}}
\def\be{\begin{equation}}
\def\ee{\end{equation}}
\def\bea{\begin{eqnarray}}
\def\eea{\end{eqnarray}}
\def\ba{\begin{array}}
\def\ea{\end{array}}
\draft
\begin{document}

\title{Nonequilibrium tube length fluctuations of entangled polymers
}
\date{\today}
\author{
G. M. Sch\"utz$^{1}$ and J.E. Santos$^{2}$}
\address{
$^1$ Institut f\"ur Festk\"orperforschung, Forschungszentrum J\"ulich,
D-52425 J\"ulich, Germany\\
$^2$ Physik Department, TU M\"unchen, James-Franck-Strasse, 85747 Garching,
Germany
}
\maketitle
\begin{abstract}
We investigate the nonequilibrium tube length fluctuations during the
relaxation of an initially stretched, entangled polymer chain. The 
time-dependent variance $\sigma^2$ of the tube length follows in the
early-time regime a simple universal power law $\sigma^2 = A \sqrt{t}$
originating in the diffusive motion of the polymer segments. The amplitude 
$A$ is calculated  analytically both from standard reptation theory and from 
an exactly solvable lattice gas model for reptation and its dependence
on the initial and equilibrium tube length respectively is discussed. The
non-universality suggests the measurement of the fluctuations (e.g. using
flourescence microscopy) as a test for reptation models. 
\end{abstract}
\baselineskip 0.3in
The most successful framework for the theoretical understanding of systems of
entangled polymers is the reptation theory developed by de Gennes, Edwards and
Doi \cite{deGe71,Doi86}. One imagines the large-scale motion of an entangled
flexible polymer as being confined to a tube which, loosely speaking, is the
sequence of pores occupied by the polymer chain within the entanglement
``network'', made up by the surrounding polymer chains. The motion of the
polymer transverse to the  tube direction is strongly suppressed due to the
topological constraints imposed by the surrounding polymers. However, mass
transport along the tube is possible by diffusion of ``defects'', i.e.,
folded polymer segments stored within the pores. The result of these dynamics 
is a snakelike random motion where polymer segments may retract into the 
existing tube at its ends (hence effectively shortening it) and subsequently 
diffuse into a new neighboring pore (hence stretching the tube and changing 
the tube contour at the end), while the bulk of the tube does not change its
shape. Because of the defects the fluctuating equilibrium tube
length $\Lambda^\ast$ is less than the actual polymer length $M$ of a fully
stretched chain. 

The basic notions of this picture have been verified
experimentally, see \cite{Rich90} for the measurement of the entanglement
distance $\delta$ implied in the tube concept, and \cite{Kaes94,Perk94}
for the role of the surrounding chains as topological constraints. Also
extensive molecular dynamics simulations confirm the reptation model
\cite{Krem90}. The theoretical status of some dynamical consequences, however,
remains rather unsatisfactory, a well-known example being the scaling
behavior of the viscosity $\eta\propto M^\kappa$ for long flexible chains. 
Doi has suggested that the discrepancy between the empirical dependence
$\kappa \approx 3.4$ and the reptation result $\kappa=3.0$ be due to
equilibrium tube length  fluctuations \cite{Doi81}, but this explanation has
not remained unchallenged \cite{desC84,Need84}. To this one may add that also
the mass dependence of the effective exponent $\kappa$ obtained by Doi appears
to be rather severe \cite{Pawi}. Since within the Doi picture such
fluctuations alter also other observable quantities through a modification of
relaxation times (in particular, the tube renewal time) we wish to address the
question to which extent standard reptation theory with underlying Rouse
dynamics for individual polymer segments provides a proper description of tube
length fluctuations. Inspired by the fact that with present experimental
techniques it is possible to observe and measure the {\em nonequilibrium} tube
relaxation process for single entangled polymer chains \cite{Perk94}, we go 
beyond the older equilibrium analysis of Refs. \cite{desC84,Need84} and
calculate here the time-dependent variance of the tube length under
far-from-equilibrium initial conditions mimicking those prepared
experimentally.

The direct observation of the reptation dynamics was achieved
in a series of remarkable experiments by Perkins et al. who have studied
entanglement effects in a dense solution of long $\lambda$-phage DNA strands
using flourescence microscopy \cite{Perk94}. Attaching a polystyrene bead to
one end of flourescence-marked DNA molecule (entangled in a background of
unmarked DNA) and pulling the bead with optical tweezers the DNA was brought
into a nonequilibrium conformation with an elongated contour directly visible 
under a microscope. Since this contour describes the  coarse-grained
path of the polymer chain it may be identified with the tube. After stopping 
the bead the contour started to contract along its own path (i.e., the tube)
as predicted by the reptation model. Since the bead was larger than the
estimated entanglement distance it could not move and the  contraction occured
only at the opposite free end of the DNA chain. Since the DNA chain used
in the experiment was very long (about 160 times its Kuhn length) it may
be regarded as sufficiently flexible to use reptation theory as an
approximation to study its tube dynamics. In
\cite{Schu99} we showed that within an universal initial time regime (but
after a very short transient relaxation process) the measured tube length
contraction is in agreement with the predictions from  both standard reptation
theory and an exactly solvable lattice gas model for reptation, related to the
Verdier-Stockmeyer model \cite{Verd62} and the repton model of Rubinstein
\cite{Rubi87}.

In order to calculate fluctuations of the tube length from reptation
theory we recall that within the underlying Rouse theory a polymer chain
with bond length $b$ is assumed to consist of $N=M/b$ frictional units, called
Rouse segments, which are connected by  harmonic springs with mean square
separation $b^2$. By not considering the actual position of the polymer in the
network but focussing only on the contour length $\Lambda$ of the tube the
reptation dynamics may be cast in a Langevin equation \cite{Doi80}
\bel{1}
\zeta \frac{\partial}{\partial t} s(x,t) = 
\frac{3k_BT}{b^2} \frac{\partial^2}{\partial x^2} s(x,t) + f(x,t)
\ee
for the position $s(x,t)$ of the Rouse segments in the tube. Here $\zeta$ is
the  friction constant of the Rouse segments and $T$ is the temperature.
Randomness is accounted for by a delta-correlated Gaussian random force
$f(x,t)$ with zero mean and $\exval{f(x,t)f(x',t')} = 
2 \zeta kT \delta(x-x')\delta(t-t')$. By definition the mean tube length is
given  by $\Lambda(t) = \exval{s(N,t) - s(0,t)}$. In the experimental set-up
considered above the position of one end is kept fixed. Hence we have to 
solve (\ref{1}) with boundary conditions $s(0,t) = 0$ and the entropic
tensile force $s'(x,t) = f_e$ at $x=N$ for all time $t>0$. The entanglement
distance $\delta$ enters only indirectly in so far as $M \gg \delta \gg b$ is
assumed. In the absence of noise (\ref{1}) has the equilibrium solution
$s^\ast(x) = x f_e$. This yields the mean equilibrium tube length
$\Lambda^\ast=N f_e$.

In order to study the tube length dynamics
it is convenient to consider $\tilde{s}(x,t) = s(x,t) - s^\ast(x)$ and
to introduce the inverse time scale $w=3k_BT/(\zeta b^2)$.
Solving first for the homogeneous equation with $f=0$ and then incorporating
the special solution of the inhomogemeous equation we obtain the full solution
of the initial value problem 
\bea
\tilde{s}(x,t) & = & \sum_{n=-\infty}^{\infty} \mbox{e}^{-wp^2_nt} 
\sin{p_nx} \times \nonumber\\
 \label{2} & & \left[\tilde{u}_n(0) + \frac{1}{\zeta} \int_0^t d\tau 
\ g_n(\tau) \mbox{e}^{ap^2_n\tau}\right]
\eea
with the Rouse ``momenta'' $p_n = \pi(2n+1)/(2N)$, the sine
transform  $\tilde{u}_n(0)=1/N\int_0^Ndy\ \sin{(p_ny)} \tilde{s}(y,0)$ of
the initial data and the sine transform $g_n(\tau)$ of the noise
term $f(x,\tau)$. For homogeneous initial stretching with initial tube length
$\Lambda_0$ one has $\tilde{s}(y,0)=(\Lambda_0 - \Lambda^\ast)y/N$ and 
obtains the mean non-equilibrium tube length 
\be 
\label{3}
\Lambda(t) = \Lambda^\ast+ (\Lambda_0 - \Lambda^\ast)\frac{8}{\pi^2}
\sum_{n=0}^{\infty} \frac{\mbox{e}^{-(2n+1)^2t/(4\tau_R)}}{(2n+1)^2}.
\ee
with the Rouse time $\tau_R = N^2/(w\pi^2)$. At early times 
$t \ll \tau_R$ one finds a universal power law behavior
$\Lambda(t) = \Lambda_0 - 2(\Lambda_0 - \Lambda^\ast)/\pi^{3/2}
\sqrt{t/\tau_R}$ \cite{Schu99}. The origin of the scaling exponent 1/2 is the
diffusive motion of the polymer segments. The reduced relaxation function
$\Delta_1(t) = (\Lambda_0-\Lambda(t))/(\Lambda_0 - \Lambda^\ast)$
depends only on time.

The tube length fluctuations are determined by the noise term in
(\ref{2}). Specifically for the variance $\sigma^2(t) = \exval{s^2(N,t)}-
\exval{s(N,t)}^2$ one finds
\be 
\label{4}
\sigma^2(t) = \frac{Nb^2}{3}\left(1-\frac{8}{\pi^2} \sum_{n=0}^{\infty}
\frac{\mbox{e}^{-(2n+1)^2t/(2\tau_R)}}{(2n+1)^2}\right). 
\ee
In the prefactor one recognizes the well-known reptation result 
${\sigma^\ast}^2=Nb^2/3$ in equilibrium \cite{Doi86}. We point out
that the amplitude depends neither on the initial stretching, i.e. on
$\Lambda_0$, nor on the equilibrium tube length $\Lambda^\ast$.
The term in the  brackets yields the growth of fluctuations due to the
randomness of the time evolution. At early times 
\be 
\label{5}
\sigma^2(t) = {\sigma^\ast}^2\frac{2\sqrt{2}}{\pi^{3/2}}\sqrt{\frac{t}{\tau_R}}
\ee
which - not surprisingly - exhibits the same universal power law dependence
as the tube length relaxation. Defining in analogy to the reduced
relaxation function the quantity 
$\Delta_2(t) =  \sigma^2(t)/{\sigma^\ast}^2$
we note the relation $\Delta_2(t) = \Delta_1(2t)$ and, for early times,
the simple relaxation ratio
\be 
\label{6}
\Delta_2(t)/\Delta_1(t) = \sqrt{2}
\ee
which does not contain any parameters of the Rouse-based reptation theory.
This feature is characteristic for universal amplitude ratios not uncommon
in scaling theory, but - as shown below - here it is an artefact of the
reptation model.

The result (\ref{3}) for the tube length relaxation has been recovered
within a lattice gas approach in which the reptation dynamics are described 
by the center of mass motion of $L = M/\delta$ individual polymer segments 
(reptons) of unit length $\delta$ which we take as the experimentally
accessible mean
entanglement distance in the melt or dense solution \cite{Schu99}. The
continuous diffusive  motion within the tube is modelled by a random hopping
of these  segments between consecutive pores with an exponential waiting time 
distribution with mean $\tau_0$. This elementary time scale is the mean
first passage time of  the motion of one repton to a neighboring pore of the
tube. To describe the  interaction of the polymer
with the network and with itself we impose the constraint that the
consecutively labeled reptons may not pass each other within the tube. End
reptons may move freely to new pores but are of course not allowed to detach
from the polymer.

We assign the label `particle' to a repton that connects two neighboring
pores and the label `hole' to a repton that is fully contained in a pore,
i.e., to a ``defect'' in the language of de Gennes (Fig.~1). The
diffusion of the defects then corresponds to particle-hole exchange with rate
$1/(2\tau_0)$, i.e. to particle hopping along the chain with hard-core on-site
repulsion.  The end-point dynamics of the tube correspond to  injection and
absorption of particles at the boundaries. These dynamics are equivalent to
the well-known one-dimensional symmetric exclusion process with open
boundaries \cite{Spoh83}. In contrast to Verdier/Stockmeyer \cite{Verd62} and
Rubinstein resp. \cite{Rubi87} we treat one boundary of the particle system 
(representing the free end of the polymer chain) as
connected to a reservoir of fixed density  $\rho^\ast =  \Lambda^\ast/M$ from
which particles are injected with rate $\alpha =  \rho^\ast/(2\tau_0)$ and
absorbed with rate  $\gamma = (1-\rho^\ast)/(2\tau_0)$. The fixed end
is described by a reflecting boundary with no particle  exchange.
We stress that the treatment of noise in this approach is rather different
from the assumptions inherent in the Langevin description (\ref{1}).
In particular, because of particle interactions in the lattice gas the noise
depends of the local hole (``defect'') density and hence implicitly on 
$\rho^\ast$ which determines the equilibrium particle density.

In this mapping the mean tube length is given by $\Lambda =\delta\exval{N}$
where $N= \sum_k n_k$ is  the total number of particles on the lattice and
$n_k=0,1$ is the particle number on site $k$. Correspondingly, $\sigma^2 =
\delta^2(\exval{N^2}-\exval{N}^2)$ where averages are taken over many
realizations of the stochastic dynamics, starting from some given
nonequilibrium initial distribution of tube lengths. An initially  fully
stretched polymer chain corresponds to an initially fully occupied  lattice.
The Markov generator of this stochastic lattice gas is known as the  quantum
Hamiltonian of the spin-1/2 ferromagnetic Heisenberg chain with boundary
field. This is an integrable model which can be fully solved by a dynamical
matrix product ansatz \cite{Stin95}, similar in  spirit to the algebraic
Bethe ansatz. The defect diffusion as described by the exclusion process is
the classical analog of the quantum spin waves in  the Heisenberg ferromagnet
and one obtains expressions for expectation values $\exval{n_{k_1}(t)\dots
n_{k_r}(t)}$ in terms of sums over Bethe wave functions \cite{Schu00}. Since 
we are interested in the early time regime $t\ll\tau_R$ during which the bulk
conformation remains unchanged we may  treat both boundaries separately. The
tethered end of the chain (caught up in the entanglement) does not contribute
to the dynamics and the particle number fluctuations induced by the free end
can be calculated by taking the thermodynamic limit $L \to \infty$. The Bethe
sums then turn into integrals over pseudomomenta (the analogues of the Rouse
modes).

The calculation of the local density $\rho_k(t) = \exval{n_k(t)}$ and
hence of $\Lambda(t)$ reduces to the solution of a relatively simple random
walk  problem \cite{Schu99} and one finds $\rho_k(t) = 1-(1-\rho^\ast)
[1-\sum_{m=1}^{2k}\exp{(-t/\tau_0)} I_{k-m}(t/\tau_0)]$. Here $I_k(x)$
are the modified Bessel functions. By identifying $w=1/(2\tau_0)$ one recovers
the Rouse result (\ref{3}) in the regime $\tau_0 \ll t \ll \tau_R$.

The calculation of the
variance requires the knowledge of the two-point correlation function
$C_{k,l}(t)=\exval{n_k(t)n_l(t)} - \exval{n_k(t)}\exval{n_l(t)}$ from which
one obtains $\sigma^2 = \delta^2 [\sum_k \rho_k(t)(1-\rho_k(t)) + 
2 \sum_{k}\sum_{l>k} C_{k,l}(t)]$. Computing the correlation function
from the Bethe wave function yields after some manipulations the exact
expression 
\bea
C_{k,l}(t) & = & F_{k+l-1,k-l}(t/\tau_0) + F_{k+l,k-l+1}(t/\tau_0)
\nonumber \\
\label{7}
 &  & - F_{k-l-1,k+l}(t/\tau_0) - F_{k-l,k+l+1}(t/\tau_0)
\eea
with
\bea
F_{m,n}(x) & = & - \frac{4(1-\rho^\ast)^2}{\pi} 
\int_0^{\pi/2}d\theta \cos{\theta}\int_0^{x}dv\mbox{e}^{2v\cos^2{\theta}} 
\times \nonumber \\
\label{8}
 & & \mbox{e}^{-2x}I_m(2(x-v)\cos{\theta})
\cos{[v\sin{(2\theta)}+n\theta]}. 
\eea
Taking the Laplace transform of this expression, summing over the lattice
indices and  expressing the result in terms of elliptic integrals finally 
yields the asymptotic law ($\tau_0 \ll t \ll \tau_R$)
\be
\label{9}
\sigma^2(t) = {\sigma^\ast}^2\frac{2g(\rho^\ast)}{\pi^{3/2}}
\sqrt{\frac{t}{\tau_R}}
\ee
with the equilibrium variance 
${\sigma^\ast}^2=\delta^2L\rho^\ast(1-\rho^\ast)$ of the lattice gas model
and $g(\rho^\ast)=1+(1-\rho^\ast)(3-2\sqrt{2})/\rho^\ast$.
The details of the rather lengthy calculation will be presented
elsewhere \cite{Sant01}. From (\ref{9}) we recover the expected scaling
exponent of the growth of fluctuations in time. However, considering a
different initial conformation \cite{Sant01} reveals that unlike in the 
standard reptation approach the amplitude depends not only on equilibrium
quantities, but also  on the initial tube length. Furthermore, the relaxation
ratio of the fully stretched chain
\be 
\label{10}
\Delta_2(t)/\Delta_1(t) = g(\rho^\ast)
\ee
turns out to be a non-universal function depending on the equilibrium defect
density $1-\rho^\ast=1-\Lambda^\ast/M$ (Fig.~2).

Our main results are the establishment of the universality
of the tube length relaxation exponent, the proof of nonuniversality of the
relaxation ratio $g$, and the unexpected lack of dependence on initial
conditions of the amplitude of the variance $\sigma^2(t)$ in standard 
reptation theory. While the
lattice gas is a phenomenological model as is standard reptation theory, it is
somewhat more microscopic in nature. Unlike the bead-spring model underlying
(\ref{1}) it incorporates the impossibility of stretching the polymer chain
and therefore lacks the physically problematic parameter $b$. Hence we
believe the master equation describing the lattice gas dynamics to be more
trustworthy than its coarse-grained Langevin counterpart. This view is
confirmed by the amplitude of the tube length fluctuations which vanish in the
lattice gas description for vanishing equilibrium defect density $1-\rho^\ast$
as they  must for a chain without defects (whose motion at the boundaries is
responsible for the tube length fluctuations). In standard reptation theory
the amplitude turns out to be unrelated to the defect density which points to 
a conceptual flaw in the treatment of the fluctuations. We conclude that the 
earlier notion that a significant modification of the equilibrium 
fluctuation theory is required extends to non-equilibrium dynamics, in
particular, to the treatment of noise. Our lattice gas approach suggests that
one should incorporate the entropic tensile force in the noise amplitude (via
multiplicative noise) in order to capture the effect of chain connectivity.

We remark that the experimental data of Ref. \cite{Perk94} are not suitable 
for a quantitative comparison with our results. On a qualitative level,
however, they lend support to the small magnitude of the theoretical 
expression (\ref{9}). In similar experiments specifically
designed for the purpose of measuring tube length fluctuations, the model
dependence of this quantity may serve as a test for the reliability of
reptation models.

J. E. S. would like to thank the EU in the framework of the
Contract ERB-FMBI-CT 97-2816 and the DFG in the
framework of the Sonderforschungsbereich SFB 413/TP C6
for financial support during the different stages of this
work. He would also like to thank the Institut f\"ur Festk\"orperforschung
at FZ J\"ulich for kind hospitality.



\newpage
\noindent Figure captions:\\[5mm]
Fig. 1: Reptation dynamics in terms of the symmetric exclusion process.
The polymer conformation in the tube (of diameter $d\approx\delta$) is
divided into twelve segments of length $\delta$ occupying six pores
corresponding to the particle configuration $A0A00AA000A0$ (from left to
right).\\[5mm]
Fig. 2: Exact relaxation ratio $g(\rho^\ast)$ calculated from the
symmetric exclusion process. The constant at $g=\sqrt{2}$ is the standard
reptation result.
\newpage

\setlength{\unitlength}{1.0cm}
\begin{figure}
\label{F1}
\begin{center}
\epsfig{width=6\unitlength,
       angle =0,
      file=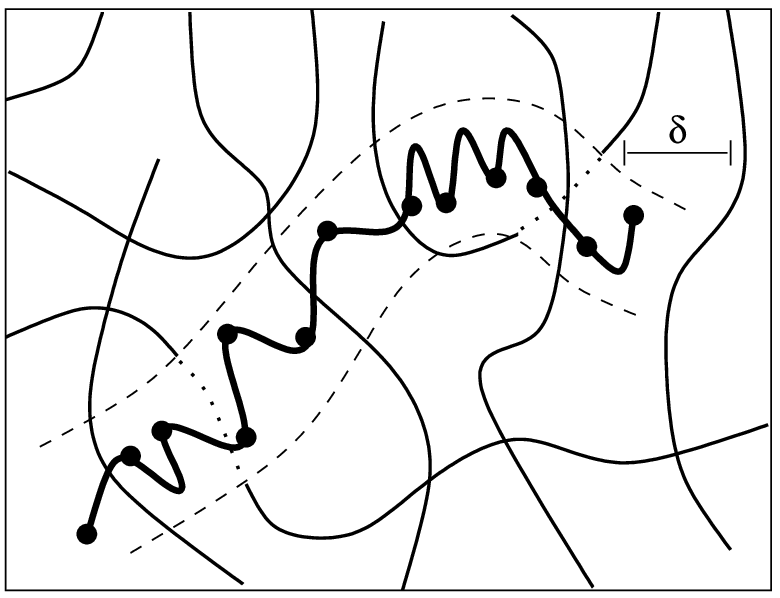}\vspace{30mm}
\end{center}
\end{figure}
Fig. 1
\newpage

\setlength{\unitlength}{1.0cm}
\begin{figure}
\vspace*{20mm}
$g(\rho^\ast)$\\[-25mm]\hspace*{1cm}
\epsfig{width=7\unitlength,
       angle =0,
      file=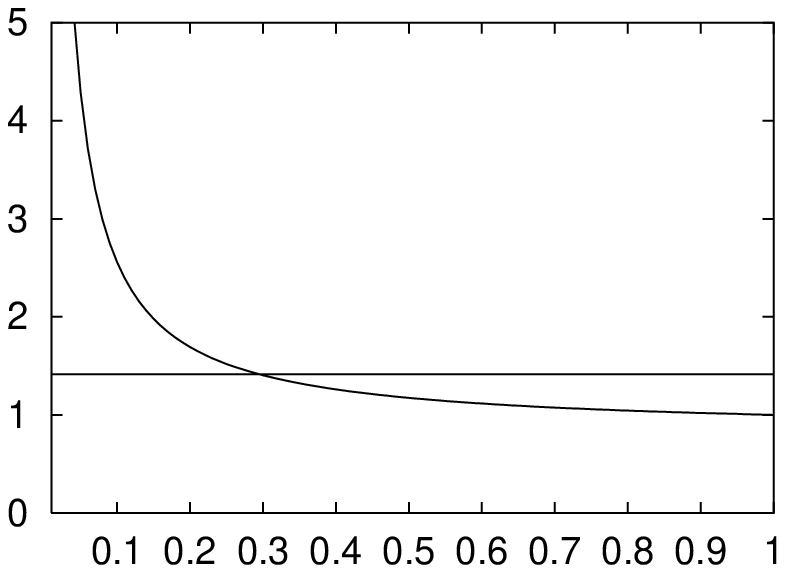}\\[-9mm]
\hspace*{80mm}$\rho^\ast$\\[30mm]
\label{F2}
\end{figure}
Fig. 2

\end{document}